\documentstyle[11pt]{article}
\newcommand{\BEQ}{\begin{equation}}  \newcommand{\EEQ}{\end{equation}}
\newcommand{\BAR}{\begin{array}}     \newcommand{\EAR}{\end{array}}
\newcommand{\BEA}{\begin{eqnarray}}  \newcommand{\EEA}{\end{eqnarray}}
\newcommand{\la}{\lambda}            \newcommand{\sig}{\sigma}

             \newcommand{\NIF}{N\rightarrow\infty}
        
\newcommand{\De}{\Delta}             \newcommand{\ra}{\rightarrow}
        \newcommand{\vph}{\varphi}

\newcommand{\hx}{\hspace*{-2mm}}     \newcommand{\GH}{{\cal G}}
\newcommand{\ha}{\frac{1}{2}}        
        \newcommand{\AM}{(A_{m-1},A_m)}
\newcommand{\ls}{\stackrel{<}{{\textstyle \sim}}} \newcommand{\al}{\alpha}
 
\newcommand{\NZ}{\hbox{I\hspace{-2pt}N}}  
\newcommand{\HH}{{\cal H}}           
\newcommand{\cy}{c=-\frac{22}{5}}    
\newcommand{\zeile}[1]{\vskip #1 \baselineskip} \newcommand{\bph}{{\bar \vph}}
  \newcommand{\CB}{{\cal
B}}
   \newcommand{\MM}{{\cal M}}
\newcommand{\rla}{\leftrightarrow}   
\newcommand{\hns}{\hspace*{-2mm}}

\newcommand{\fz}{\frac{5}{12}}       
 \newcommand{\hms}{\hspace*{-9mm}}
  \newcommand{\ny}{\nonumber}
 \newcommand{\FF}{{\cal F}}
 \newcommand{\hp}{\hspace*{6mm}}
 \newcommand{\xp}{\hspace*{3mm}}
 \newcommand{\ce}{{\tilde{c}}}
\newcommand{\df}{{\textstyle \frac{2}{5}}}  \newcommand{\JJ}{{\cal J}}
\newcommand{\ds}{{\textstyle \frac{2}{7}}} \newcommand{\ej}{\hspace*{-1mm}}
\newcommand{\ys}{{\textstyle \frac{6}{7}}} \newcommand{\xe}{{\tilde{x}}}
\newcommand{\TS}{\textstyle}               
\newcommand{\hni}{\hspace*{-3mm}}

\begin{document}
\begin{titlepage}
\null    \begin{center}   \vskip 4cm
{\huge {\bf Non-hermitian tricriticality in the Blume-Capel model with
imaginary field}  }    \vskip 1cm
{\Large  G. von Gehlen} \zeile{1}
{\large Laboratoire de Physique Th{\'e}orique, Ecole Normale Sup{\'e}rieure,
\\46, all{\'e}e d'Italie, 69364 Lyon Cedex 07, France,
 \\ and \\ Physikalisches Institut der Universit\"{a}t Bonn \\
Nussallee 12, D - 53115 Bonn, Germany} \\
\zeile{2}     \end{center} {\large {\bf Abstract :} }
Using finite-size-scaling methods, we study the quantum chain version of the
spin-$1$-Blume-Capel model coupled to an imaginary field. The aim is to
realize higher order non-unitary conformal field theories in a simple
Ising-type spin model. We find that the first ground-state level
crossing in the high-temperature phase leads to a second-order phase transition
of the Yang-Lee universality class (central charge $c=-22/5$). The Yang-Lee
transition region ends at a line of a new type of tricriticality, where the
{\em three} lowest energy levels become degenerate. The analysis of the
spectrum at two points on this line gives good evidence that this line
belongs to the universality class of the ${\cal M}_{2,7}$-conformal theory with
$c=-68/7$.
\zeile{4}   \vskip 36mm
\noindent  {\large  ENSLAPP-L-456/94~~/~~BONN-HE-94-03}
\end{titlepage}

\section{Introduction}
The study of statistical models coupled to imaginary fields can give useful
insight on the singularity structure of the partition functions and the
corresponding phase transition mechanisms. In 1952 Yang and Lee \cite{YL52}
initiated these investigations considering the Ising model in a complex
magnetic field $\CB$. In the high-temperature regime above the critical
temperature $T_c,$ the authors of \cite{YL52} found that the partition function
has zeros for purely imaginary fields $\CB= ih$ only. In the
thermodynamical limit these zeros become dense and produce a singularity
at $h = \pm h_c(T)$, which for $T \ra T_c$ gives rise to the Ising phase
transition. Fisher \cite{Fi78} pointed out that the singularity in the
$\CB$-plane discovered by Yang and Lee has many properties of a standard second
order phase transition. With the advent of the conformal field theory
(CFT)-interpretation of two-dimensional phase transitions \cite{BPZ}, Cardy
\cite{Car85} showed that the Yang-Lee singularity is described by the CFT with
the simplest possible operator content (just one primary field with conformal
dimension $d=-\frac{2}{5}$ apart from the unit operator with $d=0$)
which has the Virasoro
central charge $\cy$. This CFT is non-unitary and corresponds to
the case $p=2,~ p'=5$ of the minimal series $\MM_{p,p'}$, with central
charge \BEQ  c = 1 -6(p-p')^2/(pp').\label{cic} \EEQ
Generally, a minimal theory $\MM_{p,p'}$ ($p$ coprime $p'$) is non-unitary
if $|p-p'|>1$, since according to the Kac-formula for the scaling dimensions of
the primary fields $h_{r,s}$      \BEQ
h_{r,s}=h_{p'-r,p-s}=\frac{(rp-sp')^2-(p-p')^2}{4pp'};\hp 1\leq r\leq p'-1;
\hp 1\leq s \leq p-1,\hp r,s\in \NZ \label{Kac}\EEQ
it contains at least one primary field with negative dimension $h_{r,s}$.
If one wants to describe non-unitary minimal models by a coset construction,
Kac-Moody algebras at fractional level have to be used \cite{KacW,MaW2}.
For non-unitary minimal theories the modular invariant partition functions and
their relation to the fusion rules have been studied in \cite{ISZ,CIZ,KohS}.
Recently, such non-unitary models appeared, coupled to 2-dimensional gravity,
in the study of multicritical matrix models \cite{Kaz}. For applications of
non-unitary evolution operators in other fields of physics, see e.g.
\cite{Haa}.
\par For $p=2$ the coprime values for $p'$ are $p'=2n+3, ~n=1,2,\ldots$ (the
first member of this series is the Yang-Lee-singularity),
while for $p=3$ there are two series $p'=3n+1$ and $p'=3n+2$, etc.
So, the simplest non-unitary minimal theories beyond the $\MM_{2,5}$ are the
$\MM_{2,7}$ which has $c=-\frac{68}{7}$ and $\MM_{3,5}$ with $c=-\frac{3}{5}$.
L\"assig \cite{Lae} has considered the renormalization flow between
non-unitary theories caused by the $\phi_{(1,3)}$-perturbation. He found that
the flow leading to the $\MM_{2,5}$-theory is
$\ldots \ra \MM_{8,11} \ra \MM_{5,8} \ra \MM_{2,5}$.
So in connection with the flow to the $\MM_{2,5}$-theory it may be useful to
check the possibility of a $\MM_{5,8}$ showing up.
\par The aim of this paper is to find {\em simple} $SU(2)$-spin quantum chain
realisations of non-unitary minimal theories.
Except for $\MM_{2,5}$, which is well-known to be represented by the Yang-Lee
singularity of the Ising chain in an imaginary field, no such quantum chain
hamiltonians realizing non-unitary $\MM_{p,p'}$ theories have been written
down in the literature. There is a general, though not simple, way for
obtaining
a quantum chain model for a $\MM_{p,p'}$-theory which is even integrable:
The Forrester-Baxter \cite{FB} 2-dimensional RSOS-models have critical points
realizing $\MM_{p,p'}$-theories \cite{Ri}. Using e.g. the recipes of
DeVega \cite{HdV} one may deduce a quantum chain from the RSOS-transfer matrix.
However, this has not been done and certainly will lead to very complicated
hamiltonians. In order to get a simple realization of a higher-order
non-unitary phase transition we shall instead try to generalize the
non-integrable Ising-Yang-Lee model, looking into its spin-1 analog, and there
to possible tripel-level crossings, as will be explained below.
\par Let us briefly review the main features of the Ising representation
of the $\MM_{2,5}$-model. Already the work of \cite{YL52} and \cite{Car85}
suggested that the $\cy$ CFT should be approximated on the lattice by the
thermodynamic limit of an Ising-type quantum chain with the hamiltonian
   \BEQ  \HH_{LY}=-\ha\sum_{i=1}^N (\sig^x_i +\la\sig^z_i\sig^z_{i+1}
 +ih\sig^z_i).\label{LYi} \EEQ at a certain curve $\la(h).$    Here
$\sig^x_i$ and $\sig^z_i$ are standard Pauli matrices acting at site $i$.
$\la\sim T^{-1}$ is the inverse temperature and $h$ is the imaginary part
of the external field. For $h=0$ the model (\ref{LYi}) has the $Z_2$-symmetry
$\sig^x_i \rla -\sig^x_i$, $\sig^z_i \rla \sig^z_i$, which for $h\neq 0$
survives as an antiunitary symmetry \cite{MBD,KBD}. Due to this,
the spectrum of (\ref{LYi}) remains unchanged under the reflection $h\rla -h$.
The operator content and central charge at a particular critical point
of (\ref{LYi}) were first checked by Itzykson {\cal et al.} \cite{ISZ} and
were found to agree with the spectrum of the $\cy$ CFT.
In \cite{G} the model (\ref{LYi}) was studied with high precision by
finite-size scaling methods (FFS)
(see \cite{G} for references to earlier work which used the renormalization
group and other techniques). Fig.~1 shows the phase diagram of the model
(\ref{LYi}) as it was obtained in \cite{G}.
\par Although (\ref{LYi}) is non-hermitian, it has the special property that
the coefficients of its characteristic polynomial are real.
This is because the antiunitary property of $\HH_{LY}$ makes it a
symmetrical matrix. Therefore the eigenvalues of
(\ref{LYi}) are either real or come as complex conjugate pairs.
\par For $h=0, \HH_{LY}$ is hermitian and the spectrum is real. In the
high-temperature regime $\la<1$ there is a finite gap. By continuity, in this
regime, the finite-$N$ energy levels can become complex only if they cross and
pairwise form complex conjugate pairs. This happens first just at the $\cy$
phase transition and this property was used in \cite{G} to obtain the FFS
determination of the phase transition curve shown in Fig.~1.
For $h=0$ and $\la=1$ we have the standard Ising phase transition
described by the minimal CFT with $c=\ha$. For $h=0$ and $\la<1$ we are in the
massive $Z_2$-symmetric high-temperature phase.
\par Within a mean-field theory or Landau approach
the appearance of a phase transition for $\la<1$ and $h\neq 0$ can
be explained as follows \cite{Fi78}: Consider a model for which the
free energy admits a Landau expansion in terms of an order parameter
$\vph$ which couples to a magnetic field $\CB$:   \BEQ
\FF=\FF_0 +\CB\vph +\vph^2 \FF_2 +\vph^4 \FF_4 +\ldots   \label{LIIS}\EEQ
$\FF_0,~\FF_2$ and $\FF_4$ are functions of the inverse temperature $\la$.
For $\CB=0$ a second order phase transition will take place if there is
a $\la=\la_c$ with $\FF_2(\la_c)=0$ and $\FF_4(\la_c)>0$.
Now, for any $\la=\la_{LY}$ with $\FF_2(\la_{LY}) >0$ and $\FF_4(\la_{LY})>0$
(this is in the high-temperature phase) criticality of the system can be
restored by applying a correctly tuned {\em imaginary} magnetic field: shifting
the order parameter by the constant $i\vph_0$, i.e. $\vph\Rightarrow\bph
+i\vph_0$, (\ref{LIIS}) becomes         \BEA  \FF&=&
\CB(\bph+i\vph_0)+(\bph+i\vph_0)^2\FF_2 +(\bph+i\vph_0)^4\FF_4 +\ldots \ny\\
 &=& \ldots +\bph [\CB +2i\vph_0(\FF_2-2\vph_0\FF_4)]
 +\bph^2 (\FF_2-6\vph_0^2\FF_4)+4i\bph^3 \vph_0\FF_4 +\ldots, \label{fyl}\EEA
where the dots stand for terms of order $\bph^0,$ and higher than $\bph^3.$
The coefficient of $\bph^2$  will vanish if we choose
\BEQ \vph_0=\pm\sqrt{\FF_2(\la_{LY})/6\FF_4(\la_{LY})},\EEQ giving rise to
the Yang-Lee phase transition. This way, necessarily a term proportional
$\bph^3$ is introduced. Neglecting the residiual $\bph^4$-term,
at $\la=\la_{LY}$ we have         \BEQ   \FF= (\CB-
\CB_{LY})\bph\:+\:{\textstyle \frac{1}{3}}ig\,\bph^3+\ldots\label{phid}\EEQ
with \BEQ
\CB_{LY}=-\frac{4i}{3}\sqrt{\frac{\FF_2^3}{6\FF_4}}\hp\hp\mbox{and}\hp
  \hp g=\pm\sqrt{24\FF_2\FF_4}.    \label{phie} \EEQ
\par We now address the question, whether and how can we get non-unitary
{\em multi}critical phase transitions in the sense that not only the
coefficient
of $\bph^2$ but also higher terms of a Landau expansion come to vanish.
Since, in order to achieve this, we will need more degrees of freedom
than are available in a spin-$\ha$ chain, we shall study
a spin-1 quantum chain placed in an imaginary field which breaks the original
$Z_2$-symmetry. Is there a new type of phase transitions which may be called a
non-unitary {\em tricritical} phase transition and which is produced
by the crossing of the {\em three} lowest energy levels? In general
we should have three parameters in the hamiltonian in order to find such
a triple cross-over.
Looking into the Fisher's Landau expansion (\ref{LIIS}) this has a chance to be
realized in a model in which the parameters can be choosen to get vanishing
coefficients of $\vph^2,~ \vph^4$ and also of $\vph^3$, so that the first
nonvanishing terms is proportional $\vph^5$ (with a purely imaginary
coefficient). In a mean field treatment of the non-hermitian spin-1 model
defined in (\ref{h1}) below, this tricritical curve has been determined
by K. and M.Becker \cite{MBD,KBD}. So it seems possible to
generalize the realizations of the $\AM$ {\em unitary} minimal CFTs (which
have $|p-p'|=1$,~$p'\equiv m\ge 3$), by spin-$s$-quantum chains with
$s=m/2-1$ \cite{MBD,KBD,Dipt} to the non-unitary series. \par
Attempts to establish a connection between the mean-field non-hermitian
multicritical curve and a particular minimal CFT via a generalization of
Zamolodchikov's composite field representation of the CFT fusion rules,
have met various difficulties \cite{MBD,KBD}. So, as a more direct approach,
in this paper we treat a suitable non-hermitian quantum chain hamiltonian
by FSS methods and determine its non-hermitian tricritical manifold. From the
chain size-dependence of the low-lying levels of the spectrum we then determine
the CFT's which describe the spectrum in the different regions of the
critical manifold.
\par The main part of this paper is organized as follows: In Sec.~2 we define
the particular spin-1 quantum chain we want to study and review its
known zero magnetic field properties. In Sec.~3 we determine the
phase transition in the imaginary field from the ground-state level crossings.
Sec.~4 is devoted to the calculation of the finite-size spectrum on the
two-level crossing phase transition curve and the determination of the
universality class involved. Sec.~5 contains the main result of the paper: the
determination of the FSS of the spectrum at the tripel level crossings which
gives the evidence that there the $\MM_{2,7}$-minimal theory is realized.
Sec.~6 presents our Conclusions.

\section{The spin-1 Blume-Capel quantum chain with an imaginary field}
We consider a spin-1 quantum chain which has the property that the
characteristic polynomial of the hamiltonian matrix has only real coefficients,
so that also in this case the eigenvalues are either real or occur in complex
conjugate pairs. This property seems necessary if we want to find universality
classes described by minimal CFTs, because there the scaling dimensions of the
fields, which are related to the energy gaps of the quantum chain hamiltonian,
are always real. We insist in having a $Z_2$-symmetry for non-zero
imaginary magnetic field, because the simplest modular invariant partition
functions, those of the $p=2$-series, exhibit a $Z_2$-symmetry.
\par The non-hermitian symmetric spin-1 hamiltonian we shall consider is
defined by    \BEQ \HH = \sum_{i=1}^N\Big(\al (S_i^z)^2
    +\beta S_i^x - S_i^z S_{i+1}^z -ih S_i^z \Big). \label{h1} \EEQ
Here $S_i^x$ and $S_i^z$ are standard $3\times 3$~spin-$1$ matrices
acting at site $i$. We shall choose $S_i^z$ to be diagonal. This $\HH$ has an
antiunitary $Z_2$-symmetry: Denoting by $K$ the operator of complex
conjugation, and defining \cite{MBD,KBD}
\BEQ U=\prod_{i=1}^N \Big(2(S^x)^2-1\Big)_i,\EEQ
$\HH$ satisfies \BEQ [\HH,\Theta]=0 \hp\hp \mbox{with} \hp\hp \Theta=KU. \EEQ
There are three parameters: $\al,~\beta$ and $h$.
For $h=0$ this is the standard Blume-Capel \cite{Muk} quantum chain, and
certain linear combinations of $\al$ and $\beta$ can be interpreted as a
temperature variable and a vacancy density. In principle, one may add
more $Z_2$-symmetrical terms to (\ref{h1}), e.g. terms proportional
to $(S_i^x)^2$ and to $(S_i^z)^2(S_{i+1}^z)^2$ etc. (Blume-Emery-Griffiths
model \cite{BEG}).
However, for simplicity, here we shall consider the form (\ref{h1}) only.
\par Let us first review the phase diagram of (\ref{h1}) for $h=0$.
The critical behaviour of the Blume-Capel quantum chain (with zero field
$h$) has been studied first in mean field approximation by Gefen {\em et al.}
\cite{Muk}  and then, using finite-size scaling (FSS) by Alcaraz {\em et al.}
\cite{AlD,BDru} and by one of us \cite{Ge90}, see also \cite{Malv}. The line
$\beta=0$ is classical and is easily seen to contain a first order transition
at
$\al=1$. The spectrum of (\ref{h1}) is unchanged by the reflection
$\beta\ra -\beta$. So it is sufficient to consider only $\beta\ge 0$.
Using the FSS-technique, the transition line for $\beta \neq 0$ can
be located quite precisely as can be seen from Table ~1.    \\  \zeile{1}
\noindent \begin{tabular}{|c|cccccccccc|} \hline
$\al$&$-1.0$&$-0.5$&$ 0.0$& 0.20&0.50  &0.55  &0.60  &0.65  &0.70  &0.725\\
$\beta_c$&1.884&1.6266&1.3259&1.187&0.9411&0.895 &0.8448&0.7932&0.737 &0.707\\
\hline   $\al$ &0.76  & 0.80 & 0.84 & 0.85& 0.90 &$\underline{0.9102}$
& 0.93 & 0.95 &0.98  &1.00\\
$\beta_c$&0.6618&0.6070&0.5458&0.531&0.4365&$\underline{0.4157}$
&0.368 &0.3118&0.199 & 0.0\\   \hline  \end{tabular} \zeile{1}
Table~1: Critical values of the Blume-Capel-quantum chain, eq.(\ref{h1}) with
$h=0$, as obtained from finite-size scaling \cite{Ge90}. The underlined
values of $\al,\beta$ are those of the tricritical point. \zeile{1}
\par There is a single critical curve which starts at $\al=1$ and $\beta=0$
and then moves with increasing $\beta$ towards smaller values of $\al$.
It first follows the unit circle in the $\beta$-$\al$-plane and then takes off
to large negative $\al$, ultimately following the curve $\beta =
\sqrt{2|\al|}$.
Still close to $\al=1$, it passes a tricritical point at
\cite{Ge90}\footnote[2]{The error in the last given digit of a number will be
quoted in brackets, e.g. 0.910207(4) stands for
$0.910207\pm 0.000004$, etc.}
\BEQ \al_t=0.910207(4),\hp \xp \beta_t=0.415685(6).  \label{tric} \EEQ
For $\al <\al_t$ the phase transition is second order, and, on the
transition line, the low-lying part of the spectrum is that of the $c=\ha$
Ising CFT. The high-lying spectrum should also contain a massive component,
since the Ising Maiorana fermion needs only two of the three degrees of
freedom which are available. At the tricritical point the spectrum is
described by the $c=\frac{7}{10}$ CFT ("tricritical Ising model").
For $\alpha_t<\alpha<1$ the transition is first order. The phase below the
critical curve at $\al<1$ and $\beta\ge 0$ is a low-temperature phase with
broken $Z_2$-symmetry.

\section{Determination of the phase transition curves from lowest level
 crossings}
In analogy to the situation for the Ising chain (\ref{LYi}), we expect that
also in the high-temperature regime of (\ref{h1}) there should be
a range in $h$ where the spectrum remains real despite the hamiltonian being
non-hermitian. The spectrum should become complex where the first lowest
level crossings appear.
\par We have calculated the four lowest energy levels of the translationally
invariant states of the hamiltonian (\ref{h1}) with $N=2,\ldots,11$ sites,
for various values of the parameters $\al, \beta$ and $h$, using a Lanczos
technique. For $N\leq 8$ we diagonalize the hamiltonian also by exact methods
without the restriction to momentum-zero states.
In this paper we always take periodic boundary conditions.
\par As a typical example of the lowest level crossing behaviour at our small
values of $N$, in the last line of Table~2 we show results for
a fixed value of $\al$, i.e. $\al=0.80$, and the value $\beta=0.725$,
which for $h=0$ would be in the disordered phase, slightly above the critical
curve (for $h=0$ from Table~1 we have $\beta_c=0.6070$).

\subsection{Two-level cross-over} Now, switching on $h$, we find that as long
as
$h\ls 0.24$, the four lowest energy levels remain real as they were for $h=0$.
However, e.g. for $N=10$ sites at $h_c=0.025828$ the two lowest levels meet
and form a square-root branch point. For $h<h_c$ the two eigenvalues of
the spectrum with the lowest real parts form a complex conjugate pair. The
branch point positions are $N$-dependent.
\zeile{1}  \noindent   \begin{tabular}{|c|ccccccc|c|} \hline
\multicolumn{9}{|c|}{$\al=0.80$}  \\
$\beta$& $N=5$ & 6 & 7 & 8 & 9 & 10 & 11 & $\infty$\\  \hline
0.675&0.023796&0.019666&0.016984&0.015119&0.013758&0.012725&0.011920&0.007(1)\\
0.700&0.029975&0.025677&0.022893&0.020961&0.019553&0.018486&0.017655&0.012(1)\\
0.725&0.037034&0.032733&0.030000&0.028142&0.026815&0.025828&0.025072&0.021(1)\\
\hline  \multicolumn{9}{|c|}{$\al=0.20$}  \\
$\beta$& $N=5$ & 6 & 7 & 8 & 9 & 10 & 11 & $\infty$\\  \hline
1.800&0.143269&0.132717&0.126115&0.121739&0.118711&0.116539&0.114938&0.1080(2)\\
6.000&2.141181&2.124866&2.115361&2.109411&2.105475&2.102753&2.100700&2.0931(2)\\
\hline \end{tabular}  \zeile{1}
Table~2 : Values $h_c(N)$ of the first crossing of the two lowest levels of the
hamiltonian (\ref{h1}) for $\al=0.80$ and three values of $\beta$,
and for $\al=0.20$ and two values of $\beta$. For $h$
above these values the two lowest eigenvalues form a complex conjugate pair.
$N$ denotes the chain length, the last column $N=\infty$ gives an estimate of
branch point position in the thermodynamic limit. \\   \zeile{1}
We use standard extrapolation techniques (both rational polynomials
and Van-den-Broeck-Schwarz approximants \cite{HeS})
to obtain the branch point positions in the limit $\NIF$. In analogy to
the situation in the Yang-Lee-model (\ref{LYi})~\cite{G}, we interpret these as
second-order phase transition points.
\par  Table~3 collects our numerical results for these branch point positions
$h_c(\al,\beta)$ obtained by extrapolation $h_c=\lim_{\NIF} h_c(N)$ from
$N=3,\ldots,11$ sites. In order to save space, in Table~2 we give some
abbreviated data, but for the actual extrapolation we used also $N=3$ and
$N=4$ sites and 8-digit precision for the finite-$N$-branch point positions.
\par In Fig.~2 we plot the resulting critical surface in the space of $\al$,
$\beta$ and $h$. This surface has the shape of wings starting from the $h=0$-
critical curve of Table~1. The wings converge at a vertex which is the
standard tricritical Ising point (\ref{tric}) with $c=\frac{7}{10}$.
This shape of the wings is just the same as it is familiar for applied
staggered real fields, see e.g. Fig.~44 in Lawrie and Sarbach \cite{LauS}.

\zeile{1}  \noindent  \begin{tabular}{|ll|ll|ll|ll|ll|ll|}     \hline
\multicolumn{2}{|c}{$\al=0.20$}& \multicolumn{2}{|c}{$\al=0.50$}&
\multicolumn{2}{|c}{$\al=0.55$}& \multicolumn{2}{|c}{$\al=0.65$}&
\multicolumn{2}{|c}{$\al=0.725$}&\multicolumn{2}{|c|}{$\al=0.76$}\\
$\beta$&$h_c$& $\beta$&$h_c$& $\beta$& $h_c$& $\beta$& $h_c$&
$\beta$& $h_c$& $\beta$& $h_c$\\     \hline
1.187$\hms$&$\xp$0.0  &0.941$\hms$&$\xp$0.0  &0.895$\hms$&$\xp$0.0
&0.792$\hms$
 &$\xp$0.0  & 0.706 $\hms\hms$&$\xp$0.0 &0.662$\hms\hms$&$\xp$0.0 \\ 1.8$\hx$
&0.108(1)$\!$&1.8$\hx$&0.2172(2)$\!$&0.95$\hx$&0.0025(6)$\!$&0.85$\hx$&0.0027(3)
$\!$&0.75&0.0020(2)$\!$& 0.70&0.0018(2)$\!$\\3.0$\hx$&0.561(1)&3.0$\hx$&
0.7558(2)&1.0 &0.0070(5)&
0.95$\hx$&0.0169(8) &0.80&0.0085(3)&0.75&0.0088(3)\\  6.0$\hx$&2.08(1)
&6.0$\hx$&2.409(2)&1.1&0.0228(6)&1.0&0.0273(6)&0.85&
0.0176(5)&0.80&0.0194(3)\\9.0 $\hx$&
3.81(1)&9.0&4.21(1)&1.2&0.0445(4)&1.2&0.0820(4)&0.90&0.030(1)&0.85 &0.033(2)
\\18.&9.40(2)&18.&9.96(1)&1.8&0.246(3)&1.5&0.1922(6)&1.0
&0.059(1)&0.90&0.050(1)
\\&&&&3.0&0.807(3)&1.9&0.374(2) & & & & \\ \hline  \end{tabular} \\  \zeile{1}
Table~3 : Non-hermitian critical values $h_c(\al,\beta)$ obtained from the
lowest ground state level crossings, except for the first line, which gives
the Blume-Capel value at $h_c=0$.  \zeile{1}

\subsection{Triple cross-over: Tricritical line}
In calculating the ground-state level crossings of
the Yang-Lee model (\ref{LYi}) we don't have to worry about the third and
higher energy levels, because there always the first two levels meet before
a third level crosses in. In the spin-1-case the situation is different
because we are varying one more parameter. For fixed $0.55 <\al<\al_t$,
moving out in $\beta$, we reach a point, which we shall call $\beta_t$,
where the third energy level $E_2$ crosses the second level $E_1$ at a value of
$h$ which is lower than the crossing point of the two lowest levels $E_0$ and
$E_1$. In Fig.~3 we show this situation for $N=6$ and $\al=0.8$, where
it occurs for $\beta \approx 0.78097$. For a fixed finite chain size $N$,
there is no point where all three lowest levels meet: the tripel
crossing is an avoided cross-over since in our case there is no new
conservation law at this special point. Table~4 gives our determination of
the positions of these avoided cross-overs at chains of $N$ sites and
several values of $\al$. The calculation for obtaining these data requires
a considerable amount of computing time and therefore we mostly have not pushed
the precision beyond four significant figures, although this would be
possible if wanted. The value of $h_t$ quoted is taken somewhat arbitrarily as
the lowest value of $h$ where at $\beta=\beta_t$ we have $E_1=E_2^*$, see
Fig.~3. The convergence of the values of $\beta_t$ and $h_t$ with $N$ is quite
slow. So there is a considerable uncertainty for the values extrapolated
to $\NIF$. It may be that for $\al\ls 0.5$ both $\beta_t$ and $h_t$ move to
infinity, so that for low $\al$ the wings may extend to infinity.

\par For $\al\ra\al_t$ the values of $h_t$ go to zero. Accordingly the width
of the wings (for $\NIF$) is seen to shrink towards zero. This means that
the range in $h$ where the spectrum remains real decreases if $\al$ moves
towards the tricritical point. This is natural, since in the adjacent 1st-order
region there is a ground state degeneracy in the limit $\NIF$ at $h=0$.
\vskip 1cm   \noindent  \begin{tabular}{|c|l|llll llll|l|} \hline $\al$&$N$
 &3&4&5&6&7&8&9&10&$\infty$\\ \hline
 0.55&$\beta_t$&8.8958& 7.63& 7.45 &  7.53 & 7.66 & 7.82 & 7.887&  7.98& ?\\
 &$h_t$&{\em 4.4793}&{\em 3.61}&{\em 3.46}&
 {\em 3.49}&{\em 3.56}&{\em 3.622}&{\em 3.681}&{\em 3.73}& \\ \hline
 0.60&$\beta_t$&4.2604&3.64   & 3.462& 3.409 & 3.400& 3.4022 &3.413 &3.40&?\\
 &$h_t$&{\em 1.7276}&{\em 1.312}&{\em 1.185}&
 {\em 1.140}&{\em 1.123}&{\em 1.198}&{\em 1.121}&{\em 1.11}&  \\  \hline
 0.65&$\beta_t$&2.602&2.2336&2.1035&2.0485&2.0235 &2.0123 & 2.0075 &2.0060&
 2.006(6)\\&$h_t$&{\em 0.852}&{\em 0.613}&{\em 0.5249}&{\em 0.484}&{\em 0.464}&
   {\em 0.4535}&{\em 0.4474}&{\em 0.444}&{\em 0.437(4)} \\ \hline $\! 0.725\!$&
   $\beta_t$&1.499&1.3094&1.2310&1.191&1.1693&1.156&1.1480&1.1428&1.127(3)\\
 &$h_t$&{\em 0.356}&{\em 0.2363}&{\em 0.187}&{\em 0.161}&{\em 0.146}&{\em
0.137}
                   &{\em 0.1309}&{\em 0.1268}&0.113(2) \\   \hline
 0.76&$\beta_t$&1.204&1.062&1.005&0.9697&0.951&0.9391&0.9315&0.92596&0.908(5)\\
 &$h_t$& {\em 0.246} &{\em 0.1566}&{\em 0.119}&{\em 0.0992}&{\em 0.0872}&
           {\em 0.0796}&{\em 0.0743}&{\em 0.07062}&{\em 0.060(3)}\\ \hline

0.80&$\beta_t$&0.9482&0.849&0.806&0.780&0.7663&0.7566&0.7501&0.74524&0.728(7)\\
 &$h_t$&{\em 0.1601} &{\em 0.098}&{\em 0.071}&{\em 0.0569 }&
   {\em 0.048}&{\em 0.0419}&{\em 0.0382}&{\em 0.03529}&{\em 0.025(3)}\\ \hline
 0.84&$\beta_t$&0.748&0.6814& 0.651&0.636& 0.624 &0.617&0.6120&0.6083&?\\
 &$h_t$&{\em 0.103} &{\em 0.060}&{\em 0.0415}&{\em 0.031}&{\em 0.025}&
 {\em 0.0217}& {\em 0.0188}&{\em 0.0167}& \\   \hline    \end{tabular}
\zeile{1}
Table~4: Finite-$N$-approximations to the non-hermitian "tricritical"
points of the model (\ref{h1}). In brackets we indicate the estimated error
in the last given digit. If no bracket is given, then the error is of the order
of one unit in the last given digit. In the last column $"\infty"$ we give the
extrapolated value for $\NIF$ for those cases in which we get reasonable
convergence.
\zeile{2}
\par In Fig.~4 we show the projection of the critical curves for fixed
values of $\al$ in the $\beta - h$-plane. The curve connecting the end points
is the tricritical curve. In the mean field treatment of (\ref{h1}), which was
mentioned earlier in the Introduction, in the neighbourhood of the tricritical
point \mbox{K. and M.Becker \cite{MBD,KBD}} have obtained curves very similar
to
those of Fig.~4. They obtain the endpoints of the Yang-Lee-transition curves
through the vanishing of the Landau-coefficient of $\bph^3$ in the notation
of (\ref{fyl}), (\ref{phid}). Since in mean-field the tricritical point
for the $h=0$ model is at $(\al,\beta)=(0.5836\ldots,1.2011\ldots)$ instead
of the correct value from FSS $~(0.91021\ldots,0.41569\ldots)$, their
corresponding curves are at too small values of $\al$ and about three times
too large values of $\beta$. Anyway, it is interesting that near to the
tricritical point, mean field reproduces the shape of the fixed-$\al$
critical curves so well.

\section{Determination of the spectrum on the phase transition surface}
In order to determine the universality class of the non-hermitian phase
transitions found, we use standard finite-size-scaling (FSS) methods of
conformal theory. Using always periodic boundary conditions,
from the $N$-dependence of the ground-state energy $E_0(N)$:
\BEQ  E_0(N)/\xi=-N a_0 - \frac{\ce\pi}{6N} +\ldots  \label{cent} \EEQ  we
obtain the effective central charge $\ce$, which for non-unitary theories is
given by \BEQ \ce = c - 24 h_{min},\EEQ with $c$ the central charge and
$h_{min}=(1-(p-p')^2)/(4pp')$ is the lowest (most negative)
anomalous dimension. $a_0$ is the bulk constant which will not be of interest
in the following, $\xi$ is the conformal normalization factor \cite{CI}.
Inserting $h_{min}$ into (\ref{cic}) we can write \BEQ \ce=1-6/(pp').\EEQ
We calculate $N$ sites approximants $\ce_N$ to $\ce$ from
\BEQ \ce_N = \frac{6N(N-1)}{(2N-1)\pi\xi} \Big( (N-1)E_0(N) -NE_0(N-1) \Big),
    \label{cn}   \EEQ  and extrapolate $\NIF$.
At a conformal point the gaps scale in leading order of $N$ as $N^{-1}$. The
coefficient of $N^{-1}$ allows us to obtain the anomalous dimensions
$(h,\overline{h})$ of the primary conformal fields \cite{CI}:
\BEQ \xe_i(P)=\lim_{\NIF} \xe_i(N,P)=\lim_{\NIF} \GH_i(N,P)/\xi =(h+r+
\overline{h}+\overline{r})_i  -(h+\overline{h})_{min}  \label{dim} \EEQ
with \BEQ \GH_i(N,P) = \frac{N}{2\pi}(E_i(N,P)-E_0(N)). \label{GH}  \EEQ
Here $E_i(N,P)$ denotes the energy of the $i$-th level in the momentum
$P$-sector of the hamiltonian with $N$ sites. The positive integers $r,{\bar
r}$
 become nonzero for descendant field levels and determine the momentum
 through $P=r-r'$. The term
 \mbox{$(h+\overline{h})_{min}$} appears in case of non-unitary theories,
since there the ground state does not correspond to the vacuum.
\par The conformal normalization $\xi$ at the critical point under
consideration
is calculated from either the lowest $\De P=2$-gap in the thermal sector,
or from the lowest $\De P=1$-gap in the non-zero-charge sector \cite{CI}.
If possible, we usually use both gaps in order to have a consistency check.
\par    Tables~5 and 6 give our finite-$N$-results for the scaled gaps of
the lowest levels at two typical values of the critical surface below $h_t$
(i.e. on the wings) $(\al=0.2,~\beta=6.0,~h=2.0931)$ and
$(\al=0.5,~\beta=3.0,~h=0.7559),$  together with the values expected
for $\cy$. We see that we have excellent agreement with the $\cy$ conformal
spectrum. Our spin-1 model has one more degree of freedom than the
spin-$\ha$-Ising chain, but probably the extra degree of freedom cannot be seen
in our data because the corresponding levels are massive and are high above
the ground state.
\zeile{1}   \begin{tabular}{|c|c|ccc|cc|cc|}  \hline
\multicolumn{2}{|c|}{ }&\multicolumn{3}{c|}{$P=0$}&\multicolumn{2}{c|}{$P=1$}
& \multicolumn{2}{c|}{$P=2$} \\
$N$& $\ce$&$\xe_1$&$\xe_2$&$\xe_3$&$\xe_1$&$\xe_2$&$\xe_1$&$\xe_2$\\   \hline
2 &        & 0.312828& 0.84704& 1.2830 & 0.66008 & 0.9908 &        &        \\
3 &0.482832& 0.348901& 1.08662& 1.6274 & 0.84162 & 1.3118 & 0.8416 & 1.3118 \\
4 &0.440703& 0.367704& 1.29449& 1.8602 & 0.90864 & 1.5922 & 1.2671 & 1.7027 \\
5 &0.425037& 0.378185& 1.46168& 2.0673 & 0.93965 & 1.8434 & 1.4905 & 2.0241 \\
6 &0.417508& 0.384465& 1.58632& 2.2752 & 0.95663 & 2.0642 & 1.6166 & 2.1569 \\
7 &0.413305& 0.388459& 1.67586& 2.4853 & 0.96704 & 2.2492 & 1.6965 & 2.2195 \\
8 &0.410737& 0.391123& 1.74033& 2.6904 & 0.97395 & 2.3956 & 1.7515 & 2.2611 \\
9 &0.409083& 0.392962& 1.78775& 2.8820 &         &        &        &        \\
10&0.407955& 0.394265& 1.82350& 3.0529 &         &        &        &        \\
11&0.407336& 0.395200& 1.85109& 3.1987 &         &        &        &        \\
12&0.406839& 0.395876& 1.87279& 3.3188 &         &        & & \\   \hline
$\infty$&0.406(1)&0.41(1)&2.00(1)&3.8(1)&0.999(2)&3.01(6)&2.00(2)&2.39(4)\\
\hline $\MM_{2,5}$&0.4&0.4&2.0&4.0&1.0&3.0&2.0&2.4\\ \hline \end{tabular}
\zeile{1} \par  Table 5:~ Finite-size data for the scaled gaps $\xe(N,P)$ for
$\al=0.2,~\beta=6.0,~h=2.0931$ and the normalization $\xi=2.84$. In the bottom
line we give the values expected for the universality class is $\MM_{2,5}$.
\zeile{1}   \begin{tabular}{|c|c|ccc|ccc|c|}  \hline
\multicolumn{2}{|c|}{ }&\multicolumn{3}{c|}{$P=0$}&\multicolumn{3}{c|}{$P=1$}
& \multicolumn{1}{c|}{$P=2$} \\
$N$& $\ce$ &$\xe_1$&$\xe_2$&$\xe_3$&$\xe_1$&$\xe_2$&$\xe_3$&$\xe_1$\\ \hline
2 &        & 0.263909& 0.65712& 1.2244 & 0.66000 & 0.8367 &        &        \\
3 &0.517269& 0.307996& 0.82588& 1.5288 & 0.86748 & 1.0610 & 1.6886 & 0.8675 \\
4 &0.463568& 0.336676& 0.98033& 1.7606 & 0.94030 & 1.2531 & 2.1169 & 1.3479 \\
5 &0.440614& 0.355312& 1.12429& 1.8785 & 0.97035 & 1.4288 & 2.3901 & 1.5953 \\
6 &0.429010& 0.367693& 1.25671& 1.9671 & 0.98507 & 1.5934 & 2.6054 & 1.7750 \\
7 &0.422552& 0.376158& 1.37522& 2.0661 & 0.99321 & 1.7491 & 2.6931 & 1.9181 \\
8 &0.418705& 0.382111& 1.47783& 2.1709 & 0.99812 & 1.8964 & 2.7785 & 2.1921 \\
9 &0.416309& 0.386402& 1.56387& 2.2825 &         &        &        &        \\
10&0.414784& 0.389560& 1.63435& 2.4001 &         &        &        &        \\
11&0.413823& 0.391920& 1.69136& 2.5220 &         &        &        &        \\
12&0.413249& 0.393703& 1.73742& 2.6457 &         &        &  &  \\   \hline
$\infty$&0.38(1)&0.42(1)&2.02(3)&4.3(1)&1.011(2)&3.35(3)&3.09(6)&2.6(5)\\
\hline $\MM_{2,5}$& 0.4&0.4&2.0&4.0&1.0&3.0&2.0&2.4\\
\hline  \end{tabular}  \zeile{1}
\par  Table 6:~ Finite-size data for the scaled gaps $\xe(N,P)$ as in Table~5,
but for $\al=0.5,~\beta=3.0,~h=0.7559$ and the normalization $\xi=1.90$.
For $P=2$ the lowest levels are real for $N=3$ and $N=8$ sites, for
$N=4,\ldots,
7$ they form complex conjugate pairs. This makes the extrapolation rather
uncertain.\zeile{1}
\par As another check, whether the wings are a critical surface of the
$\cy$-universality class, we have also looked into the behaviour of the mass
gap
in the {\em neighbourhood} of the critical surface. Since the only relevant
perturbation of the $\cy$-conformal theory is by the primary field
$\phi_{(-\frac{1}{5},-\frac{1}{5})}$ (its descendent is redundant)
\cite{Car85,CMu}, we expect the mass gap to
increase proportional to $\tau^{5/12}$ where $\tau$ is the distance
from the critical curve. The particular choice of the direction in which we
take this distance is not relevant, since all directions are equivalent in
leading order, see the analogous situation in the Yang-Lee Ising model
\cite{G}. Since the perturbed $\cy$-theory contains only one single particle
of mass $m(\al,\beta,\tau)$, it follows that if the first massive level is at
$\De E_1=m$, the second massive level should appear at $\De E_2=2m$, followed
by a bunch of levels close above $2m$. Table~7 gives off-critical masses
(extrapolated to $\NIF$) for three approaches to the critical surface.
These data confirm the expected $\frac{5}{12}$-power law
and the relation $\De E_2 \approx 2\De E_1$ very well. Observe that also in
the last example, $\al=0.65, \beta=2.007, h\ra 0.431$, which is at an end
point of the critical $\cy$-surface, the data clearly show the $\frac{5}{12}$
-power behaviour. The error bars are much larger for $\De E_2$ since
for two-particle states the convergence is no longer exponential in $N$, but
only powerlike $N^{-2}$. Finite lattice effects also limit the validity of
the $\frac{5}{12}$-power law to the neighbourhood of the critical point
\cite{SaZ}.
  \zeile{1}  \renewcommand{\arraystretch}{1.3}
\noindent   \begin{tabular}{|lll|lll|lll|}   \hline
\multicolumn{3}{|c}{$\al=0.5,~\beta=3.0$} & \multicolumn{3}{|c}{$\al=0.65,
{}~\beta=1.2$} & \multicolumn{3}{|c|}{$\al=0.65,~\beta=2.007$} \\   \hline
$h$&$\De E_1$&$\De E_2$&$h$&$\De E$&$\De E_2$&$h$&$\De E_1$&$\De E_2$\\ \hline
$\ej$
0.730&$0.83208(1)\hns$&1.67(8)& 0.065& 0.3662(2)&0.73(2)& 0.35 &
0.89312(1)&1.78(1) \\  $\ej$
0.740&0.6879(1)&1.38(6)& 0.070& 0.3256(1)&0.65(3)& 0.37 &0.80688(1)&1.61(2) \\
$\ej$
0.745& 0.593(3)&1.18(4)& 0.075& 0.2687(1)&0.54(5)& 0.39 &0.69645(1)&1.36(3) \\
$\ej$
0.749& 0.494(5)&1.02(6)& 0.078& 0.2186(1)&0.44(5)& 0.41 &0.53912(1)&1.09(7) \\
$\ej$
0.752& 0.387(1)&0.81(5)& 0.079& 0.1966(1)&0.41(6)&0.415 &0.4846(3) &1.05(4) \\
$\ej$
0.753& 0.339(1)&0.67(7)& 0.080& 0.1692(1)&0.35(7)& 0.42 &0.4182(2) &0.92(5) \\
$\ej$
0.754& 0.276(6)&0.64(8)& 0.081& 0.1327(6)&0.3(1) & 0.43 &0.11(4)   &0.51(7) \\
0.7557(2)$\hni$&$\hp 0.0$& &0.08223$\hni$&$\hp 0.0$& &0.431&$\hp 0.0$ &     \\
\hline
\multicolumn{3}{|c}{$\!\De E_1\approx 2.64(2)|h-0.431|^\fz\!$}  &
       \multicolumn{3}{|c}{$\!\De E_1\approx 2.17(1)|h-0.08223|^\fz\!$} &
\multicolumn{3}{|c|}{$\!\De E_1\approx 3.97(1)|h-0.7557|^\fz\!$} \\ \hline
\end{tabular}        \renewcommand{\arraystretch}{1.0}
\zeile{1} \par Table~7: The two lowest gaps $\De E_1$ and
$\De E_2$ in the off-critical regime at various off-critical
distances converging to the respective critical points, together with, in the
bottom row, fit formulae for $\De E_1\approx \ha \De E_2$.

\section{The spectrum at "nonunitary tricriticality".}
{}From the triple crossing points determined for $N=3\ldots 10$ sites, see
Table~4, in a few cases we were able to give an estimate of the limiting values
for $\NIF$ (see the last column of Table~4). We now look into the spectra at
two
 of these values in order to see whether we can identify a particular modular
 invariant or universality class.
Taking the central values given in Table~4, we find no good evidence for
conformal invariance in the sense that for $N=2,\ldots,12,$~
the $N \De E_i$ are not clearly seen to tend to constants.
In order to improve the determination of the triple points, we did the
calculation of the spectra for a net of about ten points around the central
values found in the last column of Table~4. We looked at the respective
$N$-dependence of the effective central charge $\ce_N$ and of the $N\De E_i$
and selected the value of $\al, \beta$ and $h$ with best convergence in $N$.
This way, we were able to improve the determination of the triple crossing
points for $\NIF$ such that the $N$-behaviour required by conformal invariance
is seen.                \par
More precisely, we have calculated the spectra for $\al=0.65$ at 11 points of
the grid: $\beta=2.003;~2.007;~2.008;~2.010$ vs. $h=0.428;~0.431;~0.432;~0.438$
and choose the best convergence points. Actually, it is not worthwhile to
calculate the spectra at all points of the grid, because in the corners,
convergence is deteriorating fast, so that there we certainly move away from
the conformal point.
Analogously, for $\al=0.725$ we calculated at $\beta=0.105;~0.108;~0.110;~
0.114;~0.122$ vs. $h=1.126;~1.127;~1.128.$
Tables~8 and 9 give the finite-size data at the best convergence points.

\zeile{1} \noindent \begin{tabular}{|c|c|ccccc|ccc|}  \hline
\multicolumn{2}{|c|}{ }&\multicolumn{5}{c|}{$P=0$}&\multicolumn{3}{c|}{$P=1$}
\\
$N$&$\ce_N$&$\xe_1$&$\xe_2$&$Re\xe_{3,4}$&$\JJ_3$&$\xe_5$&$Re\xe_{1,2}$&
 $Re\xe_{3,4}$&$\xe_5$\\  \hline
2&        &0.156020&0.377460&0.79966&0.18845& 1.24913&0.4840&0.51840&1.1433\\
3& 0.44082&0.179460&0.440643&0.96901&0.19207& 1.50244&0.6305&1.14518&1.4565\\
4& 0.41810&0.197285&0.483005&1.08211&0.18728& 1.68278&0.6912&1.37159&1.6637\\
5& 0.41096&0.211759&0.513508&1.16379&0.17472& 1.84671&0.7264&1.52583&1.7996\\
6& 0.40884&0.223993&0.537305&1.22704&0.15941& 1.98653&0.7503&1.64012&1.8859\\
7& 0.40937&0.234486&0.557301&1.27863&0.14427& 2.10504&0.7685&1.73028&1.9438\\
8& 0.41190&0.243460&0.575174&1.32237&0.13036& 2.20501&0.7836&1.80464&1.9847\\
9& 0.41629&0.251001&0.591916&1.36053&0.11794& 2.28912&      &  &   \\
10&0.42261&0.257139&0.608120&1.39456&0.09705& 2.36002&      &  &   \\
11&0.43084&0.261878&0.624137&1.42542&0.09705& 2.42025&      &  &   \\
12&0.44201&0.265214&0.640163&1.45376&0.08832& 2.47204&      &  &   \\  \hline
$\infty$&?&0.30(1)&?&1.9(2)& 0.02(2)& 3.6(3) &0.97(5)&2.7(5)&2.4(2)\\
\hline$\MM_{2,7}$&0.571&0.286&0.857&$\!\!2.0/2.286\!\!$&0.0&4.0&
$\!1.0/1.286\!$&$\!3.0/3.286\!$&?\\  \hline \end{tabular}  \zeile{1}
\par Table~8:~ Finite-size spectra at the extrapolated triple crossing point
 $\al=0.65,~\beta=2.007$,\\$h=0.432$, using for the normalization factor
 the estimate $\xi=2.1$. The imaginary parts are given as
 $\JJ_i=Im E_i(N,p)/\xi$, i.e. not multiplied by the size factor $N/2\pi$.
 For $P=0$, $\xe_1, \xe_2$ and $\xe_5$ are real, but $\xe_3$ and $\xe_4$
 come as a complex conjugate pair. For $P=1$, the four lowest
 levels come in complex conjugate pairs, while $\xe_5$ is real. In the line
 $N=\infty$ we give estimates for the thermodynamic limit, which, however,
 for $\ce$ and $\xe_2$ are so uncertain that we are unable to give any reliable
 estimate. In the bottom line we list the values expected for the universality
 class $\MM_{2,7}$.
\zeile{1} \noindent \begin{tabular}{|c|c|ccccc|cc|}  \hline
\multicolumn{2}{|c|}{ }&\multicolumn{5}{c|}{$P=0$}&\multicolumn{2}{c|}{$P=1$}
\\
$N$&$\ce_N$&$\xe_1$&$\xe_2$&$\hni Re\xe_3/Re\xe_4\hni$&$\JJ_3$&$\xe_5$&
$Re\xe_{1,2}$& $Re\xe_{3,4}$\\  \hline
2&         & 0.151268& 0.372228& 0.8502/0.9530& 0.0 & 1.53302&0.5767&   \\
3& 0.568096& 0.176340& 0.445791& 0.9786/1.1829& 0.0 & 1.66967&0.7612&1.3536\\
4& 0.531168& 0.196057& 0.503739& 1.0993/1.3249& 0.0 & 1.78126&0.8595&1.6168 \\
5& 0.517239& 0.211478& 0.549269& 1.2178/1.4023& 0.0 & 1.89952&0.9040&1.7906 \\
6& 0.511390& 0.223730& 0.585726& 1.3552/1.4144& 0.0 & 2.02070&0.9323&1.9177 \\
7& 0.509887& 0.233554& 0.615615& 1.444303&$\!\!$0.07667&2.13941&0.9519&2.0166\\
8& 0.511573& 0.241395& 0.640713& 1.493708& 0.09493& 2.25193&0.9665&2.0964 \\
9& 0.516160& 0.247510& 0.662272& 1.535931& 0.10028& 2.35679&      &  \\
10&0.523733& 0.252034& 0.681178& 1.572743& 0.10003& 2.45372&      &  \\
11&0.534599& 0.255027& 0.698055& 1.605270& 0.09707& 2.54304&      &  \\
12&0.549238& 0.256493& 0.713339& 1.634270& 0.09278& 2.62529&      &  \\  \hline
$\infty$& ? &0.27(1)&0.93(6)& 2.1(2) &   ? & 3.9(2) &1.04(3)&3.1(3) \\
\hline $\MM_{2,7}$&0.571&0.286&0.857&$\!2.0/2.286\!\!\!$&0.0&4.0&
$\!1.0/1.286\!$&$\!3.0/3.286\!$ \\  \hline \end{tabular}  \zeile{1}
\par Table~9: Finite-size spectra as in Table~8, here for the extrapolated
triple crossing point  $\al=0.725,~\beta=1.126,~h=0.110$. We use the
estimated normalization factor $\xi=1.2$. Only for $N\geq 7$ we have $Re\xe_4=
Re\xe_3$. As in Table~8, we find that for $P=0$ the $\xe_1,~\xe_2$ and $\xe_5$
are real. \zeile{1}

\par We notice that there is no general tendency for the $\xe_i$ or $\ce$ to
converge towards zero, as it would appear if we missed the critical point
considerably, and instead were in the massive regime.
However, unfortunately, the sequences don't converge very well in the range
of sites ($N=2,\ldots,12$) accessible to our calculation, but one can see a
clear pattern of levels emerging.
\par       We shall now compare these patterns observed in Tables~8 and 9
to the anomalous dimensions of $\MM_{2,5}$ and three next simple minimal
non-unitary field theories $\MM_{p,p'}$ beyond the $\MM_{2,5}$: \begin{itemize}
\item $\MM_{2,7}$ with $c=-\frac{68}{7}$ and $\ce=\frac{4}{7}$,
\item $\MM_{3,5}$ with $c=-\frac{3}{5}\:$ and $\ce=\frac{3}{5}$,
\item $\MM_{5,8}$ with $c=-\frac{7}{20}$ and $\ce=\frac{17}{20}$.\\
\end{itemize} In order not to get lost in too many or somewhat
exotic possibilities, we shall not consider non-minimal models. We shall
consider only $(A_{p-1},A_{p'-1})$-modular invariants so that $x_i= 2h_i$.
This will be justified by the success of getting along with a minimal theory.
{}From the Kac-formula for the anomalous dimensions (\ref{Kac})
we calculate the conformal grids:
\renewcommand{\arraystretch}{1.4}
\zeile{1}\noindent \begin{tabular}{cccc} $\MM_{2,5}$ &
 $\MM_{2,7}$ & $\MM_{3,5}$  & $\MM_{5,8}$ \\
\begin{tabular}{|c|c|} \hline 4 & 0 \\  3 &${-\TS\frac{1}{5}}$ \\
 2&${-\TS\frac{1}{5}}$\\$ r\!=\!1$ &0\\ \hline
  & $s\!=\!1$ \\ \hline   \end{tabular} $\:$ & $\:$
 \begin{tabular}{|c|c|} \hline
6 & 0 \\  5 &${-\TS\frac{2}{7}}$ \\   4 &${-\TS\frac{3}{7}}$ \\
3 &${-\TS\frac{3}{7}}$\\ 2&${-\TS\frac{2}{7}}$\\$ r\!=\!1$ &0\\ \hline
  & $s\!=\!1$ \\ \hline   \end{tabular} $\:$ & $\:$
\begin{tabular}{|c|cc|}  \hline
4 & ${\TS\frac{3}{4}}$& 0 \\ 3 & ${\TS\frac{1}{5}}$ & ${-\TS\frac{1}{20}}$\\
2 & ${-\TS\frac{1}{20}}$& ${\TS\frac{1}{5}}$\\$r\!=\!1$&0&${\TS\frac{3}{4}}$\\
\hline                & $s\!=\!1$ & 2 \\ \hline \end{tabular} $\:$ & $\:$
\begin{tabular}{|c|cccc|} \hline      6&  ${\TS\frac{95}{32}}$&
${\TS\frac{187}{160}}$   &${\TS\frac{27}{160}}$&$-{\TS\frac{1}{32}}$\\
                      5&
${\TS\frac{7}{4}}$&${\TS\frac{9}{20}}$&-${\TS\frac{1}{20}}$&${\TS\frac{1}{4}}$\\
4&${\TS\frac{27}{32}}$&${\TS\frac{7}{160}}$&${\TS\frac{7}{160}}$&
${\TS\frac{27}{32}}$\\                                    3&
${\TS\frac{1}{4}}$&$-{\TS\frac{1}{20}}$&${\TS\frac{9}{20}}$&${\TS\frac{7}{4}}$\\
                                                          2&
$-{\TS\frac{1}{32}}$&${\TS\frac{27}{160}}$&${\TS\frac{187}{160}}$&
${\TS\frac{95}{32}}$\\
$r\!=\!1$& $0$  &${\TS\frac{7}{10}}$&${\TS\frac{11}{5}}$&${\TS\frac{9}{2}}$\\
\hline       & $s\!=\!1$    &  2   &   3   &  4    \\ \hline \end{tabular}
\end{tabular} \zeile{1}             \par Table~10: Conformal grids of the
anomalous dimensions $h_{r,s}$ for the theories $\MM_{2,5},
\MM_{2,7}, \MM_{3,5}$ and $\MM_{5,8}$ (in the grid for $\MM_{5,8}$ we have
omitted the row $s=7$ which is equal to the reversed row for $r=1$).   \\
\renewcommand{\arraystretch}{1.0}     \zeile{1}
The patterns of the spectra expected for the three cases are quite different,
and different too from the spectrum of the Yang-Lee theory $\MM_{2,5}$
which we found on the critical wings.
For the sequences $\ce;\xp\xe_1,\xp\xe_2,\xp\xe_3,\ldots$ in the
$P=0$ and $P=1$-sectors we expect (we indicate the vacuum levels by underlined
numbers):\begin{itemize} \item $\MM_{2,5}$~:  $\xp\df;\xp\underline{\df},\xp 2,
\xp 4,\xp 4+\df,\xp 6,\xp 6+\df,\xp 8,\xp \ldots$\\
$\ce= 0.4;\xp P=0~:\xp$\underline{0.4}$,\xp 2.0, \xp 4.0,
\xp 4.4,\xp 6.0,\xp 6.4,\xp 8.0,\xp\ldots$\\$P=1~: \xp 1.0, \xp 3.0, \ldots$
\item $\MM_{2,7}$~:
$\xp\frac{4}{7};\xp\ds,\xp \underline{\ys},\xp 2,
\xp 2+\ds,\xp 4, \xp 4+\ds, \xp 4+\ys,\xp 6,\xp 6+\ds,\xp\ldots$\\
$\ce\approx 0.571;\xp P=0~:\xp 0.286,\xp$\underline{0.857}$,\xp 2.0, \xp 2.286,
\xp 4.0,\xp 4.286, \xp 4.857,\xp\ldots$\\
 $P=1~: \xp 1.0,\xp 1.286, \xp 3.0,\xp 3.286,\xp \ldots$
\item $\MM_{3,5}$~:
$\xp \frac{3}{5};\xp\underline{\frac{1}{10}},\xp\ha,\xp
\frac{8}{5}, \xp 2,\xp 2+\frac{1}{2},\xp 2+\frac{8}{5},\xp\ldots$ \\
$\ce=0.6;\xp P=0~:\xp$\underline{0.1}$,\xp 0.5,\xp 1.6,\xp 2.0,\xp 2.5,
\xp 3.6,\xp 4.0,\xp 4.1,\xp\ldots $\\
$P=1~:\xp 1.0,\xp 1.5,\xp 2.6,\xp 3.0,\xp\ldots$          \item
$\MM_{5,8}$~:$\xp\frac{17}{20};\xp\frac{3}{80},\xp\underline{\frac{1}{10}},
\xp\frac{3}{16},\xp\frac{7}{16},\xp\frac{3}{5}, \xp 1,\xp
\frac{3}{2},\xp\frac{143}{80},\xp 2,\xp 2+\frac{3}{80},\xp\ldots$\\
$\ce= 0.85;\xp P=0~:\xp 0.0375,\xp$\underline{0.1}$,\xp 0.1875,
\xp 0.4375,\xp 0.6,\xp 1.0,\xp 1.5,\xp 1.7875,\xp 2.0,
\xp\ldots$\\$P=1~:\xp 1.0,\xp 1.0375,\xp 1.1875,\xp\ldots$
\end{itemize}
A first hint about which modular invariant is realized at the tripel crossing
point can be obtained by comparing $\ce$ to $\xe_1$: Both are equal for the
Yang-Lee case $\MM_{2,5}$. For $\MM_{2,7}$ we expect
$\ce=2\xe$, while we should have $\xe\ll\ce$ for $\MM_{3,5}$
and the more for $\MM_{5,8}$. Our finite-size spectra are compatible with
the second case. The normalization by $Re\xe_{1,2}^{P=1}$ is quite stable
which encourages us to look into the absolute value of $\xe_1$ and $\xe_2$.
Only for $\MM_{2,7}$ the order of magnitude is compatible with the data.
\par Also the number of levels well below $\xe_i=2$ is quite different for the
universality classes considered: one $P=0$-level in the case of $\MM_{2,5}$,
two for $\MM_{2,7}$, three for $\MM_{3,5}$ and 7 to 8 for $\MM_{5,8}$.
Since we observe two levels in this range, $\MM_{3,5}$ is not completely out
from this simple counting since the level at $\xe = 1.6$ might have been
misidentified, but we find that all data together are best compatible with
$\MM_{2,7}$.
The complex conjugate pairs in $P=0$ then must come from the first decendents
of
the two primary fields with $h=\ds$ and $h=\ys$ together. The appearance of the
imaginary parts probably is just a transitory phenomenon at small values of
$N$.
A further check of the correct assignment is given by the $P=1$-levels, which
appear at the expected positions, although again in complex conjugate pairs.
\par We see that our finite-size data strongly hint that the tripel
crossing points are described by a conformal field theory with
$c=-\frac{68}{7}$. A good further check would be calculate the spectrum of
the model with $Z_2$-twisted boundary conditions \cite{Ca,JBZ} and to look
whether it corresponds to the twisted modular invariant. It is quite
obvious to speculate that at analogous four-fold crossings in a higher spin
version of (\ref{h1}) if it contains one more parameter, the $\MM_{2,9}$-CFT
may be realized.  \par
The off-critical massive theories obtained from the $\MM_{2,2n+3}$-theories by
their $\phi_{(1,3)}$-pertur\-ba\-tions has been worked out in \cite{FKM,KM}.
The
$S$-matrix contains $N$ particles with Koeberle-Swieca-mass ratios \cite{KS},
so that the $\phi_{(1,3)}$-perturbed $\MM_{2,7}$-theory should have two
particles with mass ratio $m_2/m_1=\sin{(2\pi/5)}/\sin{(\pi/5)}$.
Since we have not tried to determine the directions of the perturbations
in the $\al, \beta, h$-plane, we cannot confirm or disprove this feature in
our spin model.

\section{Conclusions}
Using numerical finite-size scaling techniques, we have shown that the spin-1
Blume-Capel quantum chain in an imaginary field $h$, eq.(\ref{h1}), has a
two-dimensional critical surface, arising from ground-state level crossings,
which belongs to the Yang-Lee-edge $\MM_{2,5}$ $(\cy)$-universality class.
On both sides, this surface ends in a new type of tricritical line due to
tripel
lowest level crossings. Fig.~2 shows the phase diagram as determined by
finite-size scaling.    \par The FSS-spectra at the non-hermitian tricritical
lines show the particular pattern expected for the $\MM_{2,7}$-modular
invariant
partition function, from which we conclude that these realize the
$c=-\frac{68}{7}$ universality class. This is the first time that in a
simple $SU(2)$-spin quantum chain critical behaviour corresponding to a CFT
with
$c <-\frac{22}{5}$ is observed.

\subsection{Acknowledgements} The author is grateful to Katrin and Melanie
Becker for numerous fruitful and stimulating discussions on the subject.
He thanks Paul Sorba and Francois Delduc for their kind hospitality at ENS
Lyon.
This work has been supported by the DAAD
(Deutscher Akademischer Austauschdienst) though their PROCOPE-program.

\zeile{4}
\noindent {\Large\bf  Figure Captions}    \\  \\
{\bf Fig.~1:}~~Phase diagram of the Ising quantum chain in an imaginary field
$h$ defined by the hamiltonian (\ref{LYi}), from \cite{G}.   \\  \\
{\bf Fig.~2:}~~Three-dimensional view of the phase diagram of the hamiltonian
(\ref{h1}), as obtained by finite-size scaling.
The five-cornered star at $\al\approx 0.910,~\beta\approx 0.4157$
is the tricritical point with central charge $c=7/10$. It is the origin of
a pair of wings with $c=-22/5$ which extend to imaginary fields $h$ from
the Ising-like line at $h=0$ and $\al<0.910$.         \\  \\
{\bf Fig.~3:}~~Behaviour of the real part of the three lowest eigenvalues of
the spin-1-hamiltonian (\ref{h1}) near the tripel cross-over region
$\al=0.80$ and $\beta\approx 0.781$,
for $N=6$ sites (see the entry: 0.780; {\em 0.0569} in Table~4).
For $\beta<0.78098,$ $E_0$ and $E_1$ form a complex conjugate pair near
$h\approx 0.0566$. If $\beta\ge 0.78098,$ then $E_1$ joins $E_2$ to form a
complex conjugate pair above $h\approx 0.0568$. \\  \\
{\bf Fig.~4:}~~Plot of the wing critical curves obtained from FSS
as in Fig.~2, but now
for various fixed values of $\al$ projected onto the $\beta - h$ plane. \\ \\
\end{document}